\documentclass[twocolumn,natbib]{svjour3}
\usepackage{graphicx}
\usepackage{amssymb}
\usepackage{mathptmx}
\graphicspath{{figures/}}
\usepackage{subfigure}

\def\aap{\rmfamily{A\&A~}}        
\def\prd{\rmfamily{Phys.~Rev.~D~}}
\def\ssr{\rmfamily{Space~Sci.~Rev.~}}
\def\apj{\rmfamily{ApJ~}}         
\def\mnras{\rmfamily{MNRAS~}}     
\def\araa{\rmfamily{ARA\&A~}}     

\begin{document}

\title{Using ISS Telescopes for Electromagnetic Follow-up of Gravitational Wave Detections of NS-NS and NS-BH Mergers}
\titlerunning{Using ISS Telescopes for EM Follow-up of GW Detections of NS-NS and NS-BH Mergers}

\author{J.~Camp
\and S.~Barthelmy
\and L.~Blackburn
\and K.~Carpenter
\and N.~Gehrels
\and J.~Kanner
\and F.~E.~Marshall
\and J.~L.~Racusin
\and T.~Sakamoto}

\institute{Astrophysics Science Division, NASA Goddard Space Flight Center, Greenbelt, MD}


\date{Received: date / Accepted: date}

\maketitle

\begin{abstract}
The International Space Station offers a unique platform for rapid and
inexpensive deployment of space telescopes.  A scientific opportunity of great
potential later this decade is the use of telescopes for the electromagnetic
follow-up of ground-based gravitational wave detections of neutron star and
black hole mergers. We describe this possibility for OpTIIX, an ISS technology
demonstration of a 1.5~m diffraction limited optical telescope assembled in
space, and ISS-Lobster, a wide-field imaging X-ray telescope now under study as
a potential NASA mission. Both telescopes will be mounted on pointing
platforms, allowing rapid positioning to the source of a gravitational wave
event. Electromagnetic follow-up rates of several per year appear likely,
offering a wealth of complementary science on the mergers of black holes and
neutron stars.
\end{abstract}

\section{Introduction}

The ISS has recently completed its construction activities and is now moving
into its period of utilization, providing space payloads with a long-term,
stable platform\footnote{http://www.nasa.gov/mission\_pages/station/research/benefits}. Its large solar arrays and Ku-band antennas provide the ISS
with power and data capabilities which are capable of supporting multiple
payloads. The result is that scientific payloads can be attached to the ISS
without needing their own power supply or data downlink capability, and with
reduced requirements on the attitude control system, thus significantly
reducing the cost and time required to fly an experiment in space. Also, since
payloads are launched to the station on independently scheduled rockets, they
do not need to fund their own dedicated rocket, further reducing the cost of
the mission.

One such payload is the Optical Testbed and Integration on ISS Experiment
(OpTIIX), a mission to demonstrate the capability of robotically assembling a
1.5 m diffraction limited optical telescope in space. The key long-term goal of
this technology is to enable the robotic assembly of a future 10 m+ size
telescope capable of characterizing extrasolar terrestrial planets and
searching them for signs of life. 
OpTIIX also has a number of potential science applications, including monitoring of outer
solar system planetary atmospheres, stellar population studies of nearby
star-forming regions, and fast followup of astrophysical transient
events. OpTIIX has completed a Preliminary Design Review and is now
seeking funding for the construction phase, with a possible operation
start in 2017.

Another potential payload, now in the conceptual study phase to address an
upcoming NASA Mission of Opportunity, is ISS-Lobster, a wide-field imaging
X-ray telescope. Based on glancing incidence focusing of 0.3-5 keV X-rays,
ISS-Lobster is intended to provide a simultaneous wide field of view, high
spatial resolution, and high sensitivity, to advance the field of time-domain
X-ray astronomy. Its scientific goals include the study of tidal disruptions of
stars by supermassive black holes, the observation of supernova shock
breakouts, gamma-ray bursts, and the long-term variability of active galactic
nuclei.

Both OpTIIX and ISS-Lobster are planned for mounting on platforms which will be
in nearly continuous communication with ground-based alert networks through the ISS
data link, allowing the fast pointing of the telescopes in response to an
astronomical transient. This suggests the possibility of their use in an
extremely important scientific scenario around the middle of this decade, the
electromagnetic follow-up of ground-based gravitational wave detections.
Gravitational waves are produced by the interaction of extremely dense and
massive objects including black holes (BH) and neutron stars (NS). Mergers of
these compact objects also produce electromagnetic radiation: NS-NS and/or
NS-BH mergers are likely to be the source of short Gamma-Ray Bursts, a very
bright transient source of gamma, optical and X-ray signals, and also the
source of a predicted long-lived optical afterglow known as a kilonova. The
observation of BH and NS mergers is likely later this decade with the
availability of a network of gravitational wave detectors; the coincident
observation of a compact object merger event in the electromagnetic bands would
provide a rich complement to the science of the merger, and crucially, would
also help to verify the event itself. 

In this article we give an overview of the OpTIIX and ISS-Lobster designs. We
then describe the network of planned gravitational wave detectors, and then
discuss the possibilities for electromagnetic follow-up observations with the
ISS telescopes.

\section{ISS telescopes}

\subsection{OpTIIX}

\begin{figure}
	\includegraphics[width=\columnwidth]{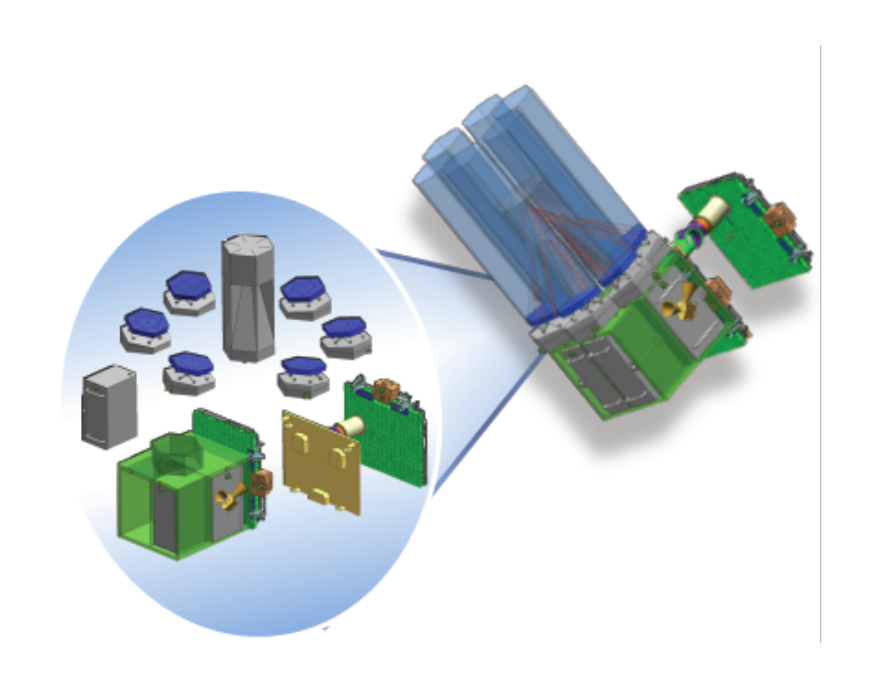}
	\caption{Exploded view of OpTIIX}
	\label{fig:1}
\end{figure}

The OpTIIX (Optical Testbed and Integration on ISS Experiment) instrument \citep{2}
consists of six modules that will be launched separately and assembled
robotically on the ISS, including six deformable primary mirror segments. An
overview of the experiment is as follows. The modules as shown in figure \ref{fig:1} are: the
Telescope Core Module, including the avionics, star trackers to determine the
telescope position, the wavefront sensing and control unit used to focus the
telescope, and the imaging camera; the Gimbal Module to provide three
degree of
freedom pointing; the Secondary Tower Module which contains the secondary and
tertiary mirrors and coarse steering mirror; and three Mirror Segment Modules,
each of which contain two primary mirror segments and their actuators, along
with laser metrology components.

The modules will be assembled on the ISS with the use of the Special Purpose
Dexterous Manipulator (SPDM), the ISS robot arm. All of the telescope
components handled by SPDM will contain the necessary interfaces and attachment
mechanisms. The SPDM has an articulating body, an assortment of tools, cameras
and sensors, and a pair of seven-jointed arms to allow efficient assembly of
OpTIIX. The ISS location of OpTIIX has not yet been determined.

The assembled telescope can be described as follows. The primary mirror,
comprised of six hexagonal segments of 50 cm diameter, has the shape of a conic
asphere, and has a diameter of 1.45 m. The primary shape is adjusted by means
of six rigid-body actuators, and 90 shape actuators, attached to each mirror
segment backing structure. The actuators provide primary figure control, as
well as high precision alignment of the segments to the telescope. Laser
metrology, enabled with a frequency stabilized laser source \citep{3,4} and beam
launchers on the primary and fiducials on the secondary, provides the
measurement of the distance and alignment between the primary and secondary \citep{5}.
The telescope is focused with image-based wavefront sensing and control
algorithms. The focus process proceeds in three stages: Shack-Hartmann sensing
for the segments initial alignment, dispersed fringe sensing for segment
co-phasing, and phase-retrieval sensing for the final wavefront adjustment to
achieve diffraction-limited performance. An imaging camera is used to obtain
the diffraction-limited images, and will include a filter wheel with eight
positions, including three wide-band and three narrow-band optical filters. The
final OpTIIX performance specifications are: diffraction limited performance
with a Strehl ratio of 80\% at 600 nm, signal-to-noise (SNR) of 5 for a star of AB
magnitude 22 or fainter in a total exposure time of 20 min, and field of view
of $\sim$3 arc min square.

Jitter noise on the ISS from anthropogenic sources as well as the expression of
ISS structural modes is attenuated with the use of a pointing platform which
also serves to position the telescope at a given sky location. This is a key
technical challenge for OpTIIX since the full telescope performance corresponds
to a sub arc sec pointing requirement. The current pointing control design
includes three loops: star trackers and gyro input to the gimbal, coarse steering mirror control
using data from the gyro, and tertiary mirror control using feedback from the
fine guidance sensor. Trade studies are ongoing.

Once assembled and operational, OpTIIX will be tasked with the acquisition of
at least 50 color images of diffraction-limited astronomical targets for
education and public outreach (EPO) use. A set of science observations will
also be undertaken, to the extent that time permits.

\subsection{ISS-Lobster}

ISS-Lobster is a wide-field X-ray telescope now under study at Goddard
Space Flight Center, planned to be situated on the ISS Express
Logistics Carrier.  It consists of two instruments, the soft
  X-ray (0.3-5 keV) Wide Field Imager (WFI), and the hard X-ray
  (10-1000 keV) Gamma-ray Transient Monitor (GTM).  It simultaneously
provides wide field of view (FOV), high position resolution, and high
sensitivity. These characteristics enable the search for transient
X-ray events.

\begin{figure}
\begin{center}
	\subfigure[Lobster focusing optic]{
	\includegraphics[width=0.8\columnwidth]{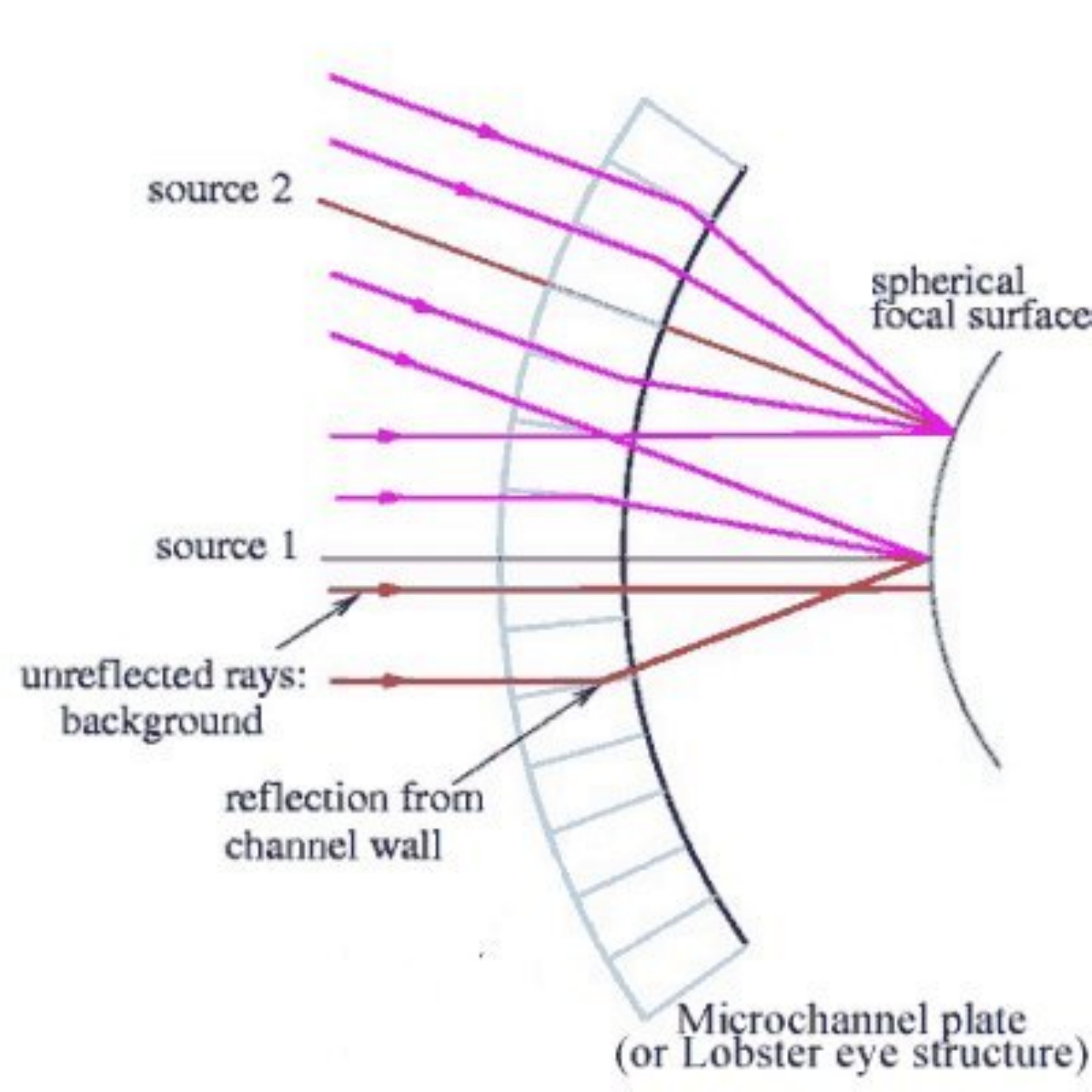}
	\label{fig:2a}}
	\subfigure[ISS-Lobster instrument on 3-axis gimbal]{
	\includegraphics[width=0.8\columnwidth]{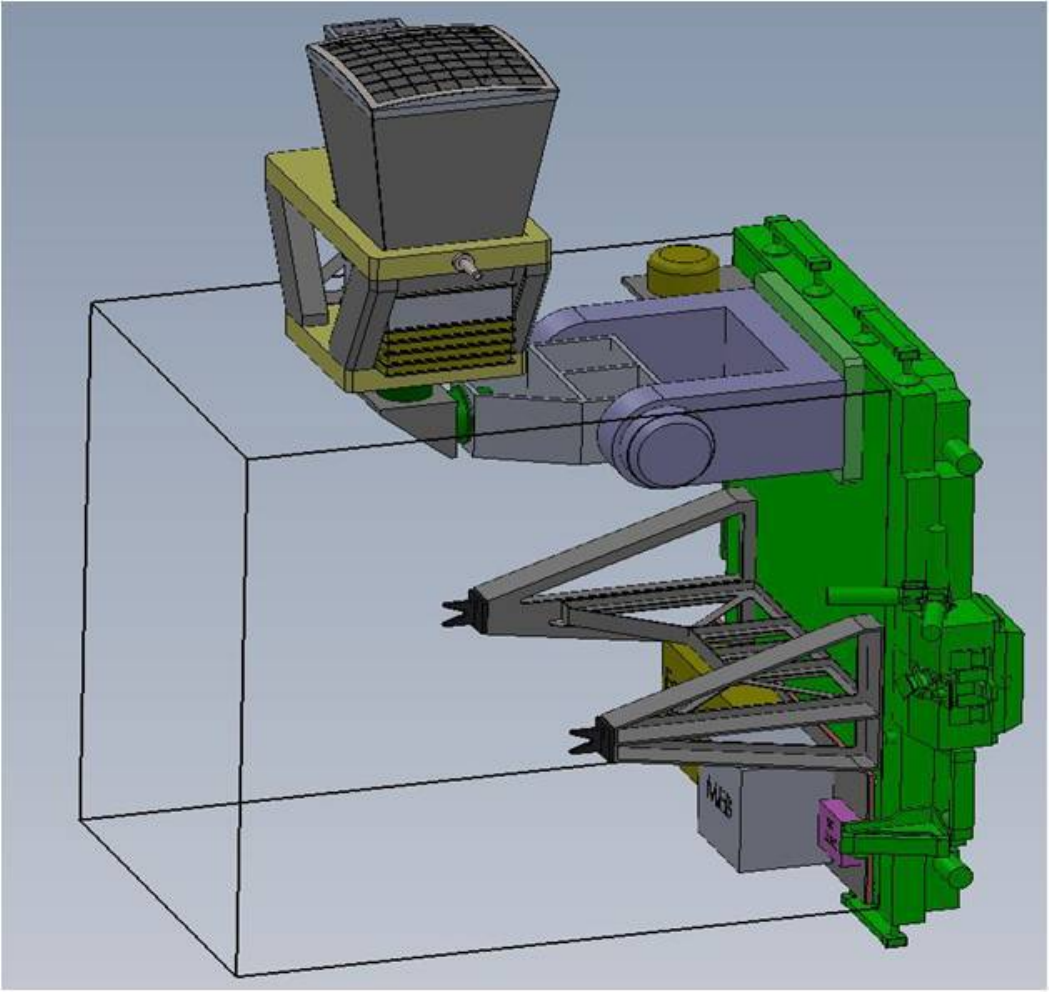}
	\label{fig:2b}}
	\caption{ISS-Lobster elements}
\end{center}
\end{figure}

Figure \ref{fig:2a} shows the principle of operation, which is similar to that employed
(in the optical band) by the namesake crustacean \citep{6}. The WFI X-ray optic consists of
an array of square-packed channels, molded into a spherical shape. The long
axes of the channels intersect a distance R along the axis of symmetry. X-rays
undergoing odd numbers of grazing reflection are brought to a focus a distance
R/2 along the symmetry axis. This gives rise to a cruciform point-spread
function consisting of a cross-shaped pattern (caused by X-ray reflections an
odd number of times in one plane and an even number of times in the other) with
a central focus, and a uniform background (caused by an even number of
reflections in both planes). The ratio of the components is about 25\% in the
focus, 50\% in the cross-arms, and 25\% diffuse background.

ISS-Lobster WFI will consist of a microchannel plate array with 75 cm radius of
curvature, with an area of 40 cm $\times$ 40 cm. The focal plane detector will be a CCD
camera, with image resolution of about 250 $\mu$m (FWHM), and energy
resolution better than 100 eV. These parameters give rise to a wide field X-ray
telescope with FOV of 30$^\circ$ x 30$^\circ$, an angular resolution
of 8 arcmin FWHM and 1 arcmin localization accuracy,
and a sensitivity of 1.3$\times$10$^{-11}$ erg/(cm$^2$\,sec) in 2000 sec for an energy range
of 0.3 to 5 keV. The focusing capability gives ISS-Lobster over a factor of 30
more sensitivity than previous wide-field X-ray telescopes -- e.g. the BAT
detector on {\it Swift} \citep{7} and the GSC detector on MAXI \citep{8}. 

The WFI module will be deployed on a pointing platform to remove the ISS
jitter, and to allow fast positioning (Figure \ref{fig:2b}) This will require a control
system consisting of a three-axis gimbal and star tracker. Initial studies have
shown that pointing at the level of 1 arcmin is achievable. 

The WFI FOV of $\sim$2\% of the sky will enable the detection of gamma-ray
bursts ($\gtrsim$0.3/day), supernova shock breakouts ($\sim$1-2/yr), and tidal disruptions of
stars by supermassive black holes ($\gtrsim$14/yr). Many other sources will be available
for study, including active galactic nuclei, stellar flares, and neutron star
bursts. 

ISS-Lobster GTM consists of a single Sodium Iodide (NaI)
  scintillation detector identical to those comprising the {\it
    Fermi}-GBM.  It will provide high time resolution data on transient
  sources over an energy range of 10-1000 keV.  While it will provide no source
  localization within its FOV, the simultaneous gamma-ray trigger with
  a GW event will confirm the astrophysical origin of the GW source, and allow for
  the possible verification of sub-threshold GW triggers.

\section{LIGO and the ground-based gravitational-wave network}

Gravitational radiation arises in regions of strong and dynamical gravitational
fields from astronomical sources, including very dense and
massive objects such as black holes and neutron stars. Its character describes
the nature of gravity in the extreme, testing the predictions of Einstein’s
theory of gravity and providing information about its sources unobtainable
through other means. In contrast to the electromagnetic waves of conventional
astronomy, which arise from the incoherent superposition of emission from the
acceleration of individual electric charges, gravitational waves result from
coherent, bulk motions of matter. Also, because gravitational waves interact
only weakly with matter they are able to penetrate the very densely
concentrated matter that produces them. The direct detection and study of
gravitational waves has the potential to revolutionize our understanding of the
universe.

A gravitational wave produces a differential strain (ratio of change in
distance to distance between two points) in space-time along orthogonal
directions, and can be observed through the relative timing of the passage of
light waves along these directions. Because gravity is a very weak force, the
strain expected at the earth from the gravitational waves of the most likely
astrophysical sources is very small, of order 10$^{-21}$. 

\begin{figure}
\begin{center}
	\includegraphics[width=\columnwidth]{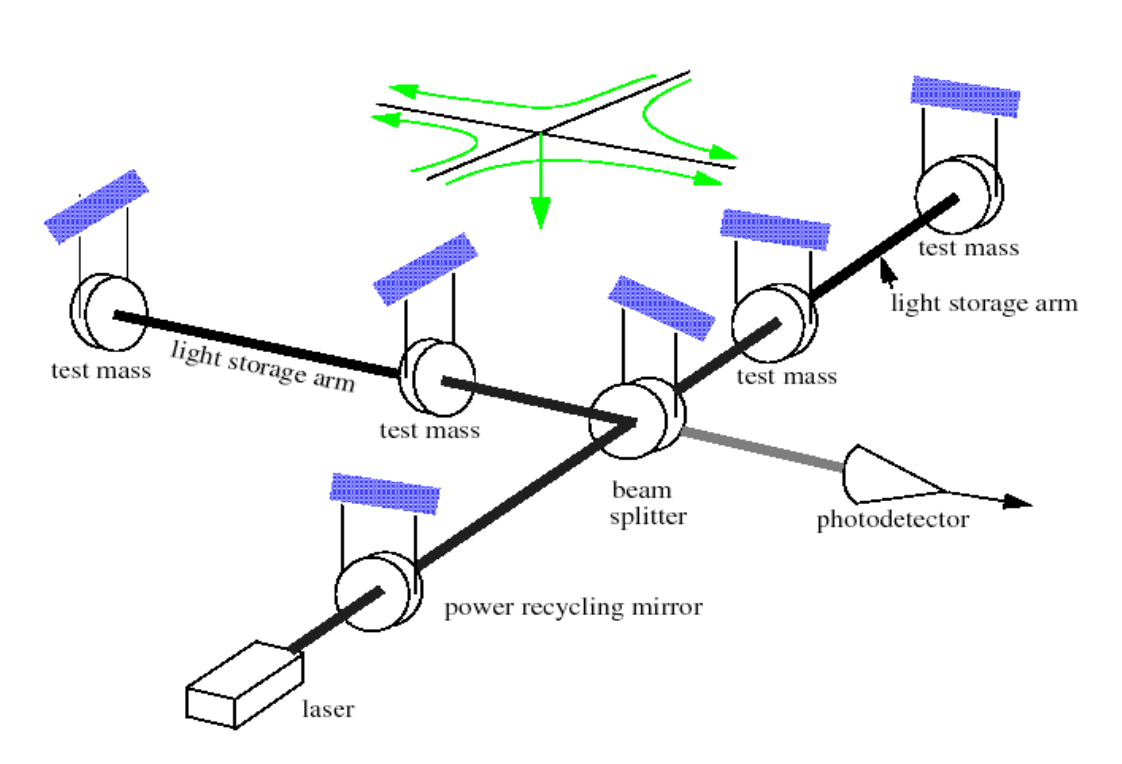}
	\caption{Gravitational wave interferometric detector schematic}
	\label{fig:3}
\end{center}
\end{figure}

A number of detectors around the world have been built, or are being
constructed, to search for gravitational waves. The general design is shown in
figure \ref{fig:3}, and is based on the principle of laser interferometry \citep{9}. A frequency
stabilized laser injects light into an interferometer beam splitter, where the
light is split and directed downstream through a light storage arm to test mass
end mirrors, and then reflected back and recombined at the beamsplitter. The
interferometer is operated so that destructive interference of the recombined
light occurs; the photodiode monitoring the output light of the interferometer
is then highly sensitive to any differential motion of the mirrors which
disturbs the exact cancellation of the light. A very high level of displacement
sensitivity, of order 10$^{-18}$ m rms, is achieved through careful attention to the
frequency and amplitude stability of the laser light, the losses of the
interferometer optics, the seismic isolation system which decouples the
interferometer from motions of the earth, and the suspension system which
stabilizes and positions the optics. Kilometer scale detectors are required to
enable sufficient strain sensitivity for gravitational wave detection.

Gravitational wave signals anticipated by these ground-based detectors include:
continuous waves from non-axisymmetric pulsars, stochastic backgrounds of
sources from the early universe, burst signals from a non-axisymmetric
supernova explosion or other unmodeled event, and signals from the inspiral and
merger of a compact binary system with a black hole and/or neutron star. The
most promising of these sources for the first detection is the NS-NS or NS-BH
merger, which becomes likely with a detector strain sensitivity of 10$^{-22}$ rms,
integrated over the  detector frequency range of 10 Hz to 2000 Hz. 

\begin{figure}
	\begin{center}
	\subfigure[Aerial view of LIGO site at Hanford, WA]{
	\includegraphics[width=0.9\columnwidth]{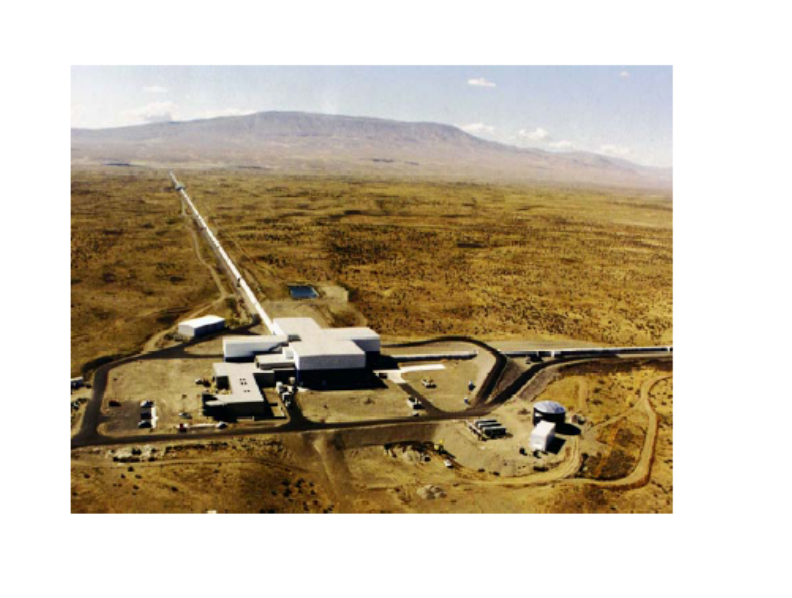}
	\label{fig:4a}}
	\subfigure[World network of GW detectors]{
	\includegraphics[width=\columnwidth]{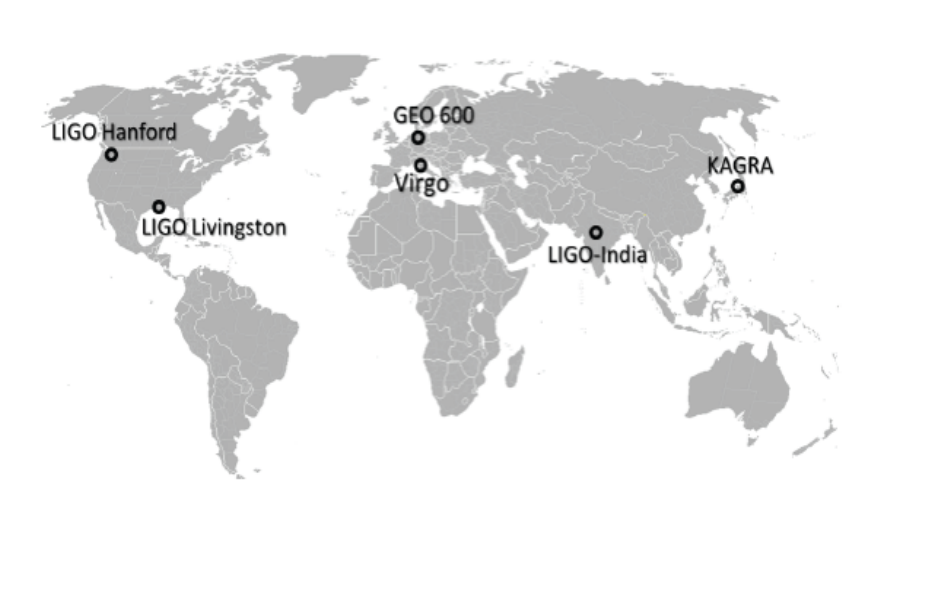}
	\label{fig:4b}}
	\caption{Current and future ground-based interferometric GW detectors}
\end{center}
\end{figure}

The gravitational wave network of detectors includes the Laser Interferometer
Gravitational Wave Observatory (LIGO) \citep{10}, with 4 km detectors in Hanford,
Washington (Figure \ref{fig:4a}) and Livingston, Louisiana; the 3 km Virgo \citep{11} observatory
in Pisa, Italy; the 600 m GEO \citep{12} interferometer in Germany; the 3 km KAGRA \citep{13}
detector under construction in Japan; and a 4 km detector recently approved for
construction in India \citep{14}. LIGO and Virgo completed in 2010 a one year period of
observation, and both projects are undergoing an upgrade to their
detectors\footnote{Advanced Virgo documentation and notes https://wwwcascina.virgo.infn.it/advirgo/docs.html} \citep{15}. Full sensitivity for these detectors is expected for LIGO and
Virgo around 2018, for KAGRA around 2020, and for the India detector in 2022.
At full sensitivity LIGO will have a detection range for a NS-NS binary merger
of 200 Mpc, and for a NS-BH merger of 400 Mpc. The detection rate for these
sources is uncertain to a factor of 10 in both directions, but the highest
probability is 40/yr and 30/yr, respectively \citep{17}.

As more detectors become operational in the network the result will be a higher
detection sensitivity, as well as a better sky location of sources: for three
detectors, the sky error box for the binary NS merger will be $\sim$100 deg$^2$ at a
network SNR of 10, but with four and five detectors the error becomes 20 deg$^2$
and 5 deg$^2$, respectively \citep{18}. For the time frame of the deployment of the ISS
telescopes, 2016--2017, the gravitational wave network will consist of the three
detectors from LIGO and Virgo.

\section{Electromagnetic Follow-up of Gravitational Wave Detections with OpTIIX and ISS-Lobster}

\subsection{Electromagnetic counterpart science}

The detection of electromagnetic counterparts to gravitational wave signals
from these NS-NS and NS-BH mergers with OpTIIX and/or ISS-Lobster will
powerfully enhance the science of the compact merger event. In brief, the
detection of an EM counterpart will:
\begin{itemize}
	\item{Locate the galaxy of the source progenitor. The position resolution
			of OpTIIX and ISS-Lobster is at the level of 1 arcsec and 1 arcmin,
			respectively, both easily high enough to locate the progenitor
			galaxy for an event within the LIGO detection range of 400 Mpc.}
	\item{Determine the source redshift. With the galaxy location determined,
			ground based spectroscopic telescopes can follow-up the detection
			and determine the host redshift. When combined with the source
			distance that can be extracted from the gravitational waveform,
			this could lead to a more precise local determination of H$_0$, the
			Hubble constant \citep{19}.}
	\item{Deliver information on source beaming,
			and examine the afterglow energetics to probe the progenitor and
			its gas environment.}
	\item{Test predictions of General Relativity,
			including the speed and polarizations of gravitational waves.}
	\item{Lift degeneracies associated with the binary source parameters \citep{20}.
			Extraction of key binary source information such as luminosity
			distance and inclination is limited by the poor sky localization of
			the Gravitational Wave (GW) network. The precise sky localization enabled by the EM
			counterpart signal will assist in the analysis of the gravitational
			waveform, allowing significantly higher precision in the extraction
			of source parameters.}
	\item{Raise the confidence level of a GW
			detection. The coincident measurement of an EM signal with a GW
			detection, both spatially and temporally, will greatly aid in
			reducing false detections. This can allow the lowering of the GW
			detection threshold, extending the range of the measurement.}
	\item{Determine the engine of the short Gamma-Ray Burst (sGRB). The sGRB,
			extensively studied by the NASA mission {\it Swift} \citep{21}, is believed to be
			due to a compact object merger. Analysis of a gravitational wave
			signal coincident with an EM signal from an sGRB will definitively
			determine whether the event is due to an NS-NS or NS-BH merger.}
\end{itemize}

In the ISS telescope observational scenarios described next, we will consider
first the known optical and X-ray afterglow signatures which follow the prompt
gamma-ray signal of the sGRB. The sGRB afterglows decay
with a several hour timescale, and are excellent candidates for follow-up
detection within an ISS orbital period of 90 minutes. The engine for the
extremely high energy sGRB has been established with a reasonable degree of
confidence as an NS-NS or NS-BH merger by examining the stellar populations
associated with observed sGRBs \citep{22}; this implies a known gravitational
wave signal associated with this source. We can estimate the rate of a GW/X-ray
or GW/optical coincidence for the sGRB as follows. The observed sGRB rate
\citep{23,32} is $\sim$10/(Gpc$^3$\,yr) (which includes beaming of the prompt
gamma-ray into a nominal 15 deg cone); within the advanced LIGO range of 200
Mpc for a randomly oriented NS-NS merger, this gives an afterglow coincidence
rate of 0.3/yr, assuming the afterglow has the same beaming as the prompt
signal. However two factors enhance this rate. First, the amplitude of a
gravitational wave is peaked along the GRB axis by a factor of 1.5. Second, the
requirements of a time coincidence between the GW and prompt gamma-ray, and a
spatial coincidence between the GW and the X-ray or optical afterglow source
locations, can lower the GW network threshold by a factor of 1.5 while keeping
the same false positive rate\footnote{The sensitivity improvement from
	requiring an EM coincidence depends on both the gravitational-wave
	background rejection factor, and the empirical shape of the background
	about the threshold of interest. The factor of 1.5 used in this calculation
	is consistent with a factor of 10$^7$ GW background rejection from the
	requirement of a prompt short gamma-ray coincidence within 5 sec and X-ray
	afterglow within 100 deg$^2$, and a GW background distributed approximately
	exponentially with rate $\sim$ 100$^{-\mathrm{SNR}}$ about SNR=10 (e.g. the
	two-detector NS/NS background in \cite{34}, figure 3. The X-ray transient
	rate within 100 deg$^2$ was estimated from \cite{35}, which looked for
	transients visible for 7.5 hours or less at a flux limit of
	$\sim$10$^{-12}$ erg/s/cm$^2$. 32 potential candidates were found in a
	survey coverage of 76,000 deg$^2$ day, giving a rate of $\sim$10$^{-4}$
	transients per deg$^2$. This may be extrapolated to 10$^{-3}$ per 100
	deg$^2$ at 10$^{-11}$ erg/s/cm$^2$ as a rough estimate for the Lobster
	transient rate.}. Together these two factors will extend the gravitational
wave detection range by 2.3, leading to a coincidence rate of 4/yr. For the
NS-BH merger engine, the advanced LIGO range is 400 Mpc, and similar reasoning
gives an EM/GW coincidence rate of 30/yr.  These rates need to be adjusted for
the efficiency of the X-ray and optical afterglow detection by the ISS
telescopes, discussed below.

We also consider the signature for a so-called kilonova \citep{25}, a predicted (but not
yet observed) optical afterglow from a NS-NS merger, powered by heating from
the radioactive decay of heavy elements created in the merger ejecta. A
significant advantage in the detectability of the kilonova coincidence with a
gravitational wave is that its optical emission is believed to be isotropic, so
that its rate is not limited by the possibility of beaming away from the
earth \citep{26}. Since all NS-NS mergers are expected to result in a kilonova, we use a
NS-NS merger rate of 1,000/(Gpc$^3$\,yr) \citep{17} and a sky-average range of 200 Mpc for
randomly oriented NS-NS mergers to predict a GW-EM rate of 40/yr. (For the
kilonova there is no preferred orientation, and the optical afterglow is
expected to occur amidst a significant  background level, and thus the GW range
enhancements discussed above do not apply. Also, the difference in kilonova and
sGRB rates of a factor of 100 can be reconciled through the beaming associated
with the sGRB emission.) Finally, kilonova emission still occurs in low-density
environments where sGRB optical afterglows become very weak.

\subsection{ISS telescope pointing}

In the event of a gravitational wave candidate signal from an NS-NS or NS-BH
merger, an alert can be issued to the astronomical community within 5
minutes \citep{27}. The alert will contain the sky localization error box associated
with the event, and will be immediately available to the ISS telescopes through
the continuous ISS data link. Both OpTIIX and ISS-Lobster will be mounted on
pointing platforms that can slew to a given sky position, with a time lag
dependent on the ISS position. The maximum delay for ISS-Lobster to access the
LIGO-generated sky position will be slightly less than 1/2 the period of the
ISS orbit, or 40 min, with an average delay equal to half the maximum, or 20
min. OpTIIX is expected to be able to schedule follow-up observations within
the next orbit, which makes it an excellent complement to a related
ground-based optical telescope follow-up effort \citep{28,29}; a further advantage of
the OpTIIX ISS deployment is the increased angular resolution possible above
the atmosphere.

In order to maximize the likelihood of detecting the
  electromagnetic counterpart to a GW trigger, ISS-Lobster will begin
  pointed observation as soon as a GTM short duration trigger occurs
  by tiling the $2\pi$ steradian GTM FOV with the WFI in 25 short ($\sim$1
  min) pointings starting from the zenith direction, until a source is either
  detected or the GW position is uploaded.

The main impediment to full sky viewing on the ISS are the solar panels, whose
position varies on a daily and seasonal basis, and the Sun; after accounting
for these factors, the average viewing efficiency
of ISS-Lobster and OpTIIX are expected to be about 80\%. To avoid excessive
repointings from the large number of false positives associated with LIGO
sub-threshold triggers, a coincidence with the sGRB prompt gamma-ray signal,
obtained with the ISS-Lobster NaI detector, will be required before either
telescope is slewed to the LIGO sky position. 

\begin{figure}
	\includegraphics[width=\columnwidth]{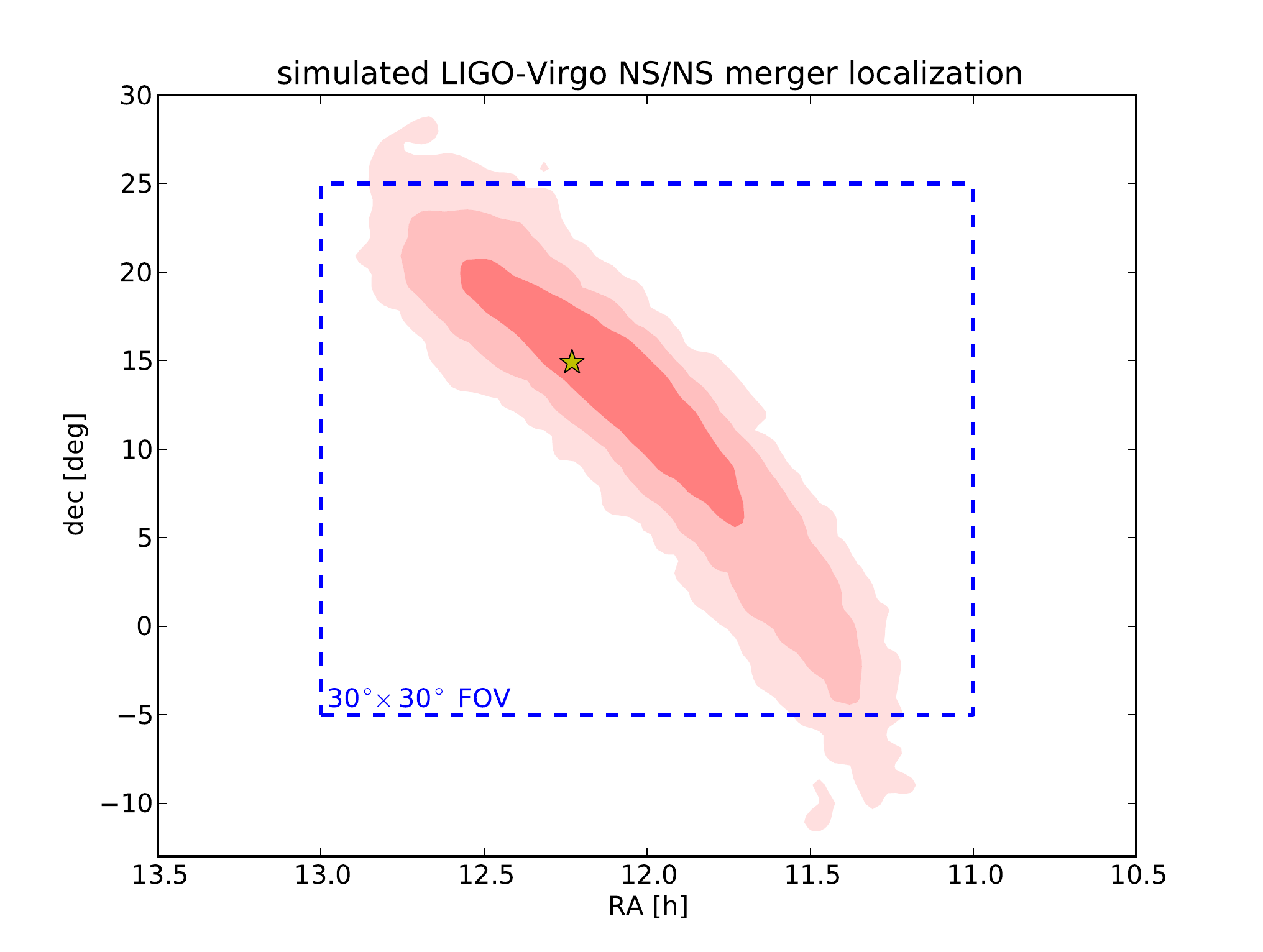}
	\caption{ISS-Lobster FOV (dotted line) and NS-NS location error box \citep{36} where contours indicate 1, 2 and 3$\sigma$ confidence. The star indicates the true source location.}
	\label{fig:5}
\end{figure}

An important issue for the telescopes is locating the source within the 100
deg$^2$ gravitational wave error box. ISS-Lobster has a significant advantage
here, as its large 30$^\circ$ $\times$ 30$^\circ$ field of view allows it to encompass
the gravitational wave sky error box from a NS-NS merger with a network SNR of
10, as shown in figure \ref{fig:5}, and localize the source to 1 arcmin. The large FOV
for ISS-Lobster is optimal for the early phase of operation of the
gravitational wave detector network ($\sim$2017), with three detectors and a sky
localization error box of 100 deg$^2$. In contrast, covering the full error box by
tiling observations with OpTIIX would require an excessively large and
impractical number of pointings, since its FOV is only 3 arcmin. Instead, the
detection of the optical afterglow of an sGRB could be enabled by first
localizing the source position with ISS-Lobster to 1 arcmin, so that it can be
contained within the OpTIIX field of view. In this case the OpTIIX detection
rate is limited by ISS-Lobster.

\begin{figure}
	\subfigure[{\it Swift} sGRB X-ray afterglows scaled to GW
          horizon (400 Mpc). The vertical lines represent the fastest
          response times for the follow-up scenarios.  At least 80\%
          of the afterglows will be detectable by ISS-Lobster.]{
	\includegraphics[width=\columnwidth]{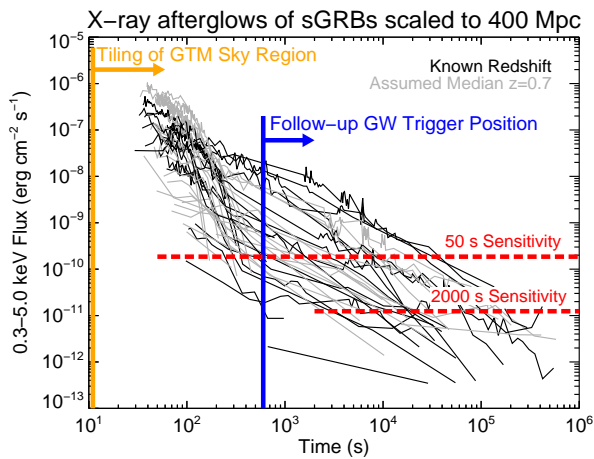}
	\label{fig:6a}}
	\subfigure[sGRB optical afterglows with OpTIIX threshold]{
	\includegraphics[width=\columnwidth]{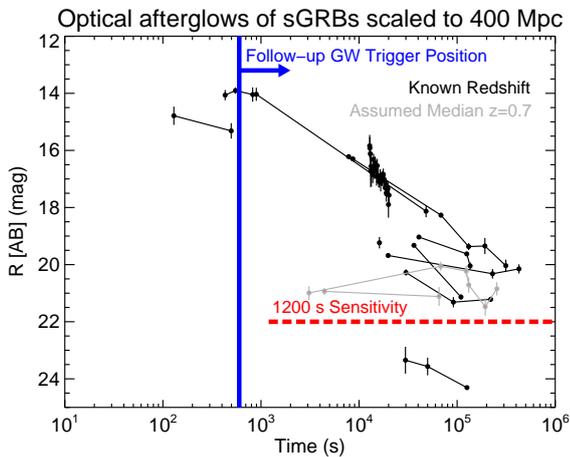}
	\label{fig:6b}}
	\caption{{\it Swift} known-sGRB afterglow light-curves with flux re-scaled to 400 Mpc}
	\label{fig:6}
\end{figure}

If ISS-Lobster is not available, another possible solution, for sources within
100 Mpc, is to restrict OpTIIX pointings to galaxies within the LIGO range,
since the location of a compact merger (as seen as an sGRB) appears to track
the total mass of galaxies. For this purpose a galaxy catalog was assembled
from publically available information, known as the GW Galaxy Catalog (GWGC) \citep{30}.
This catalog (60\% complete at present), contains $\sim$50,000 galaxies out to a
distance of 100 Mpc, with roughly one galaxy per square degree of sky, or 100
hosts for a typical sky position error of 100 deg$^2$. By using a weighting with
the galaxy luminosity (a proxy for total mass), the potential number of hosts
can be further reduced to $\sim$40, which captures 90\% of the total catalog luminosity, and
can finally be reduced to 10, by requiring the capture of 50\% of the total
luminosity, resulting in (0.6 $\times$ 0.5) or 30\% total probability of detection. This is a
number of pointings (1 or 2 per orbit) that could be realistically accomplished
by OpTIIX within the day-long decay time of an optical afterglow. An upgrade to
GWGC to increase its listings, optimize its performance and increase its range
is ongoing\footnote{J. Kanner, private communication}.

\subsection{ISS-Lobster follow-ups}

At least 80\% of sGRBs produce an X-ray afterglow, as detected
by {\it Swift} within the first few hundred seconds.  The
efficiency of detecting an X-ray afterglow with ISS-Lobster is
indicated in figure \ref{fig:6a}, where the flux from 43 sGRBs (17
with measured redshift, and 26 with no redshift) is plotted versus
time, with the redshift used to scale the source flux to a distance of
400 Mpc (those GRBs with no measured redshift are considered to be
located at a median redshift of 0.7.)  The vertical lines
  indicate the minimum times to begin WFI observation after a GTM sGRB
  trigger and subsequent tiling, or follow-up observations of GW
  trigger position.  Roughly 80\% of the afterglows will be detectable
  with the WFI. Including the ISS viewing efficiency (80\%), a
  conservative X-ray/GW coincidence rate is 1-2 sGRB/yr (at 440 Mpc;
  NS-NS) and 12 sGRB/yr (at 900 Mpc; NS-BH). These rates are also
  based on the assumption that the GW detector network is operating
  at full sensitivity (200 Mpc for LIGO, 100 Mpc for Virgo) and 80\%
  duty cycle.


We note that these rates assume that the X-ray afterglow is beamed to the same
degree as the prompt signal, however the evidence for this is weak at best, as
evidence for beaming (an achromatic break in the X-ray light curve) is seen in
only two or three of the 43 sGRB afterglows \citep{32}. If the afterglow is more
isotropic than the prompt signal, as some models assume \citep{33,37}, the X-ray/GW
coincidence rates will go up accordingly (in this case we may modify the prompt
coincidence requirement with the GTM.)

\subsection{OpTIIX follow-ups}

About 30\% of all sGRBs have optical afterglows. The optical afterglow to an
sGRB within the LIGO detection range will appear bright (16--18 mag) at times
less than 5 hours after the gravitational wave alert, and then dim
significantly over the course of a day. Thus OpTIIX will be able to observe the
decay of the light curve by repeated exposure to the source over multiple ISS
orbits. Figure \ref{fig:6b} shows the flux from 13 sGRBs (8 with measured redshift, and 5
with no redshift) plotted versus time, with the redshift used to scale the
source flux to a distance of 400 Mpc (those GRBs with no measured redshift are
considered to be located at a median redshift of 0.7.) The vertical dotted line
is set at the average time required for a source to come in view by OpTIIX, and
the horizontal line shows the sensitivity. All the afterglows are bright enough
to be detected. Assuming OpTIIX is pointed to the source location by first
imaging the X-ray afterglow to 1 arcmin with ISS-Lobster, the sGRB optical/GW
coincidence rate is 1/yr for the afterglow of the NS-NS merger engine, and 4/yr
for the NS-BH merger. Again, these rates will increase if the afterglow beaming
is more isotropic than the prompt gamma-ray.

The isotropic kilonova optical afterglow has a longer timescale than that of
the sGRB. It is predicted to peak around 1 day after the NS-NS merger with a
magnitude between 19 and 22, and decay over the following 3 days. Again
assuming that the source is first localized by the (beamed) X-ray afterglow
with ISS-Lobster, the OpTIIX detection rate for the kilonova optical/GW
coincidence is $\sim$2/yr. In contrast, if the source is within 100 Mpc, the galaxy
map can be used to guide the OpTIIX search for the kilonova signal with about
10 pointings, yielding a rate of about 1/yr given the 30\% detection
probability achievable with the galaxy catalog mentioned above. This may be an
important followup source in the time period before LIGO reaches its full
sensitivity. Finally, addressing the difficult problem of distinguishing a
kilonova from background transient events including supernovae and variable
stars will require an accurate measure of the candidate signal time variance,
measured over successive ISS orbits. The ISS deployment of OpTIIX will mean
that atmospheric fluctuations will not be present in the measurement of source
variability (a key challenge for ground-based optical telescopes).

\section{Conclusions}

The ISS platform offers a unique and inexpensive opportunity for
electromagnetic follow-up studies of gravitational wave sources. These
follow-up studies will offer a wealth of complementary science including
determining the source location and redshift, removing degeneracies in the GW
analysis, and raising the signal-to-noise of the event. The technology
demonstrator OpTIIX, a 1.5 m diffraction limited optical telescope, is
scheduled for ISS operation in 2016. ISS-Lobster, a time-domain X-ray
observatory, is in the conceptual design phase for a proposal to a Mission of
Opportunity, which could be operational in 2017. These schedules coincide well
with the planned operation of the LIGO and Virgo gravitational wave detectors.

Both OpTIIX and ISS-Lobster will be mounted on pointing platforms in continuous
communication with ground-based gravitational wave alerts, allowing the fast
acquisition of the source sky location. The source location within the large GW
sky error box ($\sim$100 deg$^2$) expected in the initial operating phase of the GW
network can be imaged within the FOV of ISS-Lobster to 1 arcmin, allowing a
follow-up by OpTIIX and localization to 1 arcsec. Use of a galaxy catalog can
also independently guide the pointing of OpTIIX if the source is within 100
Mpc.

With the assumptions that 1) a short GRB is powered by either an NS-NS or an
NS-BH merger, 2) the X-ray and optical afterglows of an sGRB are beamed into
the same opening angle as the prompt gamma-ray, and 3) a kilonova is the
correct description of the long-term optical afterglow of a NS-NS merger, the
coincident rates for EM/GW signals are: X-ray signal from sGRB with NS-NS
(NS-BH) merger engine, 2/yr (12/yr); optical signal from sGRB with NS-NS
(NS-BH) merger engine, 1/yr (4/yr); optical signal from NS-NS kilonova using
galaxy catalog to guide pointing, 1/yr; optical signal from NS-NS kilonova
using ISS-Lobster to localize source, 2/yr. These rates encourage the planned
use of these ISS telescopes for electromagnetic follow-up to gravitational wave
detections, an exciting emerging astronomical field which is certain to contain
sources and surprises beyond those described here.

\end{document}